\documentclass[a4paper]{jpconf}
\RequirePackage{fix-cm}
\usepackage{graphicx}
\usepackage{setspace}
\usepackage{graphics}
\usepackage{amsmath}
\usepackage{amssymb}
\usepackage{float}
\usepackage{epstopdf}
\newcommand{\vs}{\vspace{3mm}}

\newcommand{\be}{\begin{equation}}
\newcommand{\ee}[1]{\label{#1} \end{equation}}
\newcommand{\ba}{\begin{eqnarray}}
\newcommand{\ea}[1]{\label{#1} \end{eqnarray}}
\newcommand{\nl}{\nonumber \\}

\begin{document}

\title{
Quarks, Flow and Temperature in Spectra
}
 
\author{T.S.~Bir\'o, Z.~Szendi}
\address{
  MTA Wigner Research Centre for Physics, Budapest, Hungary 
}

\ead{biro.tamas@wigner.mta.hu, insane.zsu@gmail.com}

\author{Z.~Schram}
\address{
 Department of Theoretical Physics, University of Debrecen, Debrecen, Hungary \\
 MTA-DE Particle Physics Research Group, Debrecen, Hungary
}

\ead{schram@phys.unideb.hu}


\begin{abstract}
 Strangeness in Quark Matter 2013 Theory Review
\end{abstract}

\section{Introduction}

\vs
A good theory summary should not consist of a direct enumeration of talks, rather a critical
reflection, possibly a synthesis of different viewpoints is expected. Here we follow
this concept: after identifying the key fundamental questions in the research field
on quark matter five competing views will be presented as duelling alternatives.
The physics behind experimental findings may not only differ from, but sometimes even
contradict one another, as well as mathematical model approaches can be based on
diametrally opposite philosophies. We shall exhibit a few simple theoretical
exercises in order to illuminate these alternatives.
By its very nature this summary is subjective.

In our view there are three fundamental questions about high energy matter:
i) the degree of stohasticity in the phenomena we experience,
ii) the particle -- wave (particle gas -- strong field) duality in the 
description of the acting objects, and iii) the reliability of the 
conjectured initial state without any direct experimental evidence.
''The Lord does not play dice'' - surmised Albert Einstein, but quite often
it looks like that: we lack information even then, when physically it would
not be prohibited to measure it. We are blind to part of the phase space
by the design of our detectors (we hope only to an unimportant part).
We are trying to model and understand {\em average} outcomes from
hundreds of millions of phyiscally independent events by statistical and
continuum physics calculations and we do not have much exclusive information
event by event. Furthermore if we had, we possibly could not digest it.
We are talking about (point) particles, but calculate with (plane wave) fields.
We model initial and intermediate state fluctuations by guessing and expect to see
an emerging pattern on the statistics of outcomes.
But what are our choices when doing theoretical calculations? What are
viable alternatives? What has been our progress lately in answering these
fundamental questions? In this review we try to collect some partial answers.

\newpage
The main ''idea-wise'' duels occur as follows:
\begin{enumerate}
		\item thermalization:  exponential vs power-law $p_T$ spectra
		\item hydrodynamics:  initial state vs eos-driven flow features
		\item hadronization:  phase space vs QCD matrix element dominance
		\item Quark Matter eos:  lattice vs dual gravity methods
		\item nuclear medium:  cross sections vs mean fields
		\item simulation:  particle vs field based models
	\end{enumerate}
In the this review we plan to discuss many of these dichotomies.

Last but not least, before starting to review theoretical contributions
to this conference, let us recall Helen Caines' introductory talk about
experiments: we have gathered more information from experiments, seeing
more size, more time and more gradients in heavy ion collisions.
From the overview of several different  beam energies we learn
approximate relations between yield and rapidity range,
\be
 \frac{dN_{{\rm ch}}}{d\eta} \propto \ln \sqrt{s} \sim y_{{\rm beam}},
\ee{CAINES_YIELD}
between volume and charged particle rapidity yields
\be
V\approx 730 {\rm fm}^3  + 2.545 \frac{dN_{{\rm ch}}}{d\eta},
\ee{CAINES_VOLUME}
and between time duration and charged particles per rapidity
\be
{\cal T} \approx 0.875 \left( \frac{dN_{{\rm ch}}}{d\eta} \right)^{1/3}. 
\ee{CAINES_TIME}
These formulas are approximate fits to the figure she showed\cite{HELEN}.

\section{Thermalization}

\vs
The competition between thermal and non-thermal views of the intermediate ''fireball''
state of high-energy heavy ion collisions is reflected in the hunt for exponential
spectra. Besides that any continuous spectrum can have a straight line fit on the
logarithmic plot in a short enough interval, claiming thermal phenomena must be based
on a thermal state, on having (at least local) equilibrium in the energy distribution
of the parts of the surmised physical system. Therefore the theoretical study of 
possible ways of thermalization, of reaching such a state, has key importance.

Either studying the emergence or assuming the existence of a thermal state,
thermal and hydrodynamical models are numerous. Just citing some of the
titles from the program: 
''Thermalization of massive partons in anisotropic medium (Florkowski),
First 3 seconds (Rafelski), Thermal model (Stachel),
Strangeness balance in HIC (Kolomeitsev),
Systematic Properties of the Tsallis Distribution (Cleymans),
Thermalization through Hagedorn States (Beitel),
Relativistic distribution function... (Grossi),''
already shows the wealth and divergence among theoretical approaches.
While some say that ''the fireball never thermalized, but was born in a
thermal state'', such a statement also requires explicit, 
quantitative calculations\cite{UNRUH,SATZ,CASTERINA,BIRO}.

Since the hydrodynamical approach assumes local thermal equilibrium,
these two theoretical tools should be treated together. We summarize the
very essence of the relativistic hydro- and thermodynamics as follows.

The formulas are based on the Local Conservation of Noether Currents (LCNC),
 \be
  \partial_{\mu} J^{\mu \, a} \: = \: 0.
  \ee{HYDRO1}
Here $J^{\mu a}$ is a Lorentz four-vector indexed by $\mu$ ranging over
all spacetime directions, while $a$ stands generally for any further
collection of indices related to the independent generators of infinitesimal
symmetry transformations.
More specifically the essential  LCNC-s used in our theories are related
to conserved charges, like baryon (B), strangeness (S) or electric charge (Q);
to spacetime shifts ($\nu$) and rotations and Lorentz-boosts (hyperbolic
rotations with $\nu\rho$ antisymmetric index pairs):
\ba
\partial_{\mu} J^{\mu \, (B,S,Q)} &=& 0,
\nl \nl
\partial_{\mu} T^{\mu \, \nu} &=& 0,
\nl \nl
\partial_{\mu} M^{\mu \, \nu\rho} &=& 0.
\ea{HYDRO2}
In building up hydrodynamics it is essential to explore the
connections between these currents.
Chemistry deals with several types of charges ($q_i^a$): 
\be
J^{\mu \, a} = \sum_{i=+,-}\limits q_i^a \, u_{(i)}^{\mu}
\ee{CHEMI}
denote conductive currents, carried by the common stream, $u^{\mu}$, or by
several streams, $u_{(i)}^{\mu}$ in multicomponent fluids. 
There can be further directions in transport, too. 
Conduction or radiation\footnote{Conduction may be viewed
as phonon radiation, its speed is also limited by that of the light.}
can be directed also transverse to the matter charge flow:
This is most prominent in the energy-momentum. Its symmetric descriptive tensor can be
splitted to several terms aligned and orthogonal to a fiducial flow velocity
vector, $u^{\mu}$:
\be
T^{\mu\nu} = P^{\mu} u^{\nu} + P^{\nu} u^{\mu} + {\mathfrak{T}^{\mu\nu}}
\ee{HYDRO_T}
with $P^{\mu}=e u^{\mu}+q^{\mu}$.
Finally when polarisation related phenomena are considered then the
angular momentum and spin density tensor has to be taken into account:
\be
M^{\mu\nu\rho} = x^{\mu}T^{\nu\rho}-x^{\nu}T^{\mu\rho}+S^{\mu\nu\rho}
\ee{HYDRO_SPIN}
is anti-symmetric in all its index pairs.

Expectations and hopes by using hydordynamics in the description of quark
(and sometimes hadronic) matter include
\begin{itemize}
\item Determination of Properties of Matter (equation of state and transport coefficients),
\item Comparison with Lattice QCD,
\item Determination of Initial State just after the Collision
\item Key for the early dynamics...
\end{itemize}
By its nature the majority of our efforts in relativistic hydrodynamics 
and in the underlying kinetic theory approaches are numerical.
A characteristic example of how far such efforts can go and how strong the visualising
power of the numerical results can be was presented by Joannis Bouras
showing the evolution of Mach cones in BAMP simulations\cite{BAMPS,CUBA}.

On the other hand theoretical efforts to improve
the classical hydrodynamical calculations do not 
weaken\cite{AGUIAR,SOCOL,TAKAHASHI,OSADA,WILK,OWI,BIRO+MOLNAR,DENICOL,NIEMI,DENI2}.
In his review talk at this conference Takeshi Kodama presented a picture of a magnifying glass
revealing that in reality the flowing matter is not smooth, but very much fluctuating.
This coarse graining problem opens the hydrodynamics at its ultraviolet end,
at small sizes towards possible improvements. At the infrared end, because of missing
infinitely large reservoirs even in the heaviest ion collision, the
thermodynamical treatment has to be modified by finite size effects\cite{BIROideal,BIROBV}.

By this point we have to clarify the connection between relativistic
hydrodynamics and thermodynamics. In fact, after considering the
entropy four-current, $S^{\mu}$, thermodynamics follows from the above
described general structure of the basic system of equations.
During a generic process dissipation occurs, closed systems show a tendency
of approaching an equilibrium state. The net and local production of
entropy is non-negative:
 \be
 \partial_{\mu}S^{\mu} + \lambda_a \partial_{\mu} J^{\mu, a} \ge 0.
 \ee{ENTROPY_GROWS}
Here all LCNC-s (locally conserved Noether currents) have to be taken
into account, each with its own Lagrange multiplier, $\lambda_a$.
Classically the energy-momentum tensor plays a prominent role,
its Lagrange-multiplier,  $\lambda_{\nu}=\beta_{\nu}$, describes
a local, but moving thermometer. Its invariant length
represents a Doppler-shifted local temperature\footnote{With the relative velocity
between the local flow and the thermometer.}
\cite{BIROEPL}.

\vs
The thermodynamic Gibbs potential analogue is a four-vector, an integral
of the above (in a standing body it becomes $p/T$ the pressure over the temperature):
\be
 S^{\mu} + \lambda_a J^{\mu \, a} = \Phi^{\mu}
\ee{GIBBS}
is the first theorem of thermodynamics. From eqs.(\ref{ENTROPY_GROWS}) and
(\ref{GIBBS}) follows the
Gibbs-Duhem relation (valid in and out of local equilibrium):
\be
  \partial_{\mu}\Phi^{\mu} \ge J^{\mu \, a} \partial_{\mu} \lambda_a 
\ee{DUHEM}
Finally linear transport coefficients, $\eta_{ab}$, are defined as positive semi-definite
in order to ensure the inequality (\ref{ENTROPY_GROWS}):
 \be
 \partial^{\mu} \lambda_a = - \eta_{ab} J^{\mu \, b}.
 \ee{TRP}
In higher order hydrodynamics the evolution (relaxation) of such coefficients is also
calculated.

\vs
However, is this the truth? Do we really produce smooth, flowing and relativistic continua in
high-energy experiments? Is there also a definite, sharp-valued thermodynamic temperature?
Or we experience pseudo-thermalization and pseudo-flow?
Some hints towards that this might be the case dates back to 1975, when it was shown
that simple acceleration may produce a thermal feeling, it may lead to radiation
spectra appearing as the black-body radiation
\cite{UNRUH,SATZ,CASTERINA,BIRO}.

The physical reason is simple: a moving monochromatic source radiating with
a single frequency, $\omega$, is seen shifted according to the relativistic
Doppler-effect:
\be
\omega' = \omega \sqrt{\frac{1-v(\tau)}{1+v(\tau)}}.
\ee{REL_DOPPLER}
If the source accelerates, its momentaneous velocity, $v(\tau)$ depends on
the proper time along its trajectory. With a constant deceleration, $v(\tau)=v(0)-g\tau$.
The spectral analysis Fourier transforms the ever changing phase by the
changing Doppler factor obtaining an intensity distribution according to
\be
I(\Omega) \propto \left|\int\!e^{i\int\omega'(\tau')d\tau'}e^{-i\Omega\tau} \, d\tau\right|^2.
\ee{INTENS}
This intensity divided by $\hbar\Omega$ delivers a photon number yield per
invariant phase space element; observing only the high energy part of spectra
also massive (but light) particles might show similar effects.
Finally it is just a mathematical fact that the integral in eq.(\ref{INTENS})
is proportional to\footnote{It is derived by using a new integration variable, $z$,
which satisfies $\omega'(\tau)=dz/d\tau$.}
\be
I(\Omega) \propto \frac{1}{e^{2\pi \Omega / g}-1}.
\ee{UNRUH}
By this $T=g/2\pi$ is the Unruh temperature in Planck units.
For a velocity change in the order of $1$ (lightspeed) in a time traversing
about $1$ fm, it is really in the order of $150$ MeV. By this we are talking
about extreme accelerations.

\begin{figure}[H]
\includegraphics[width=12pc,angle=-90]{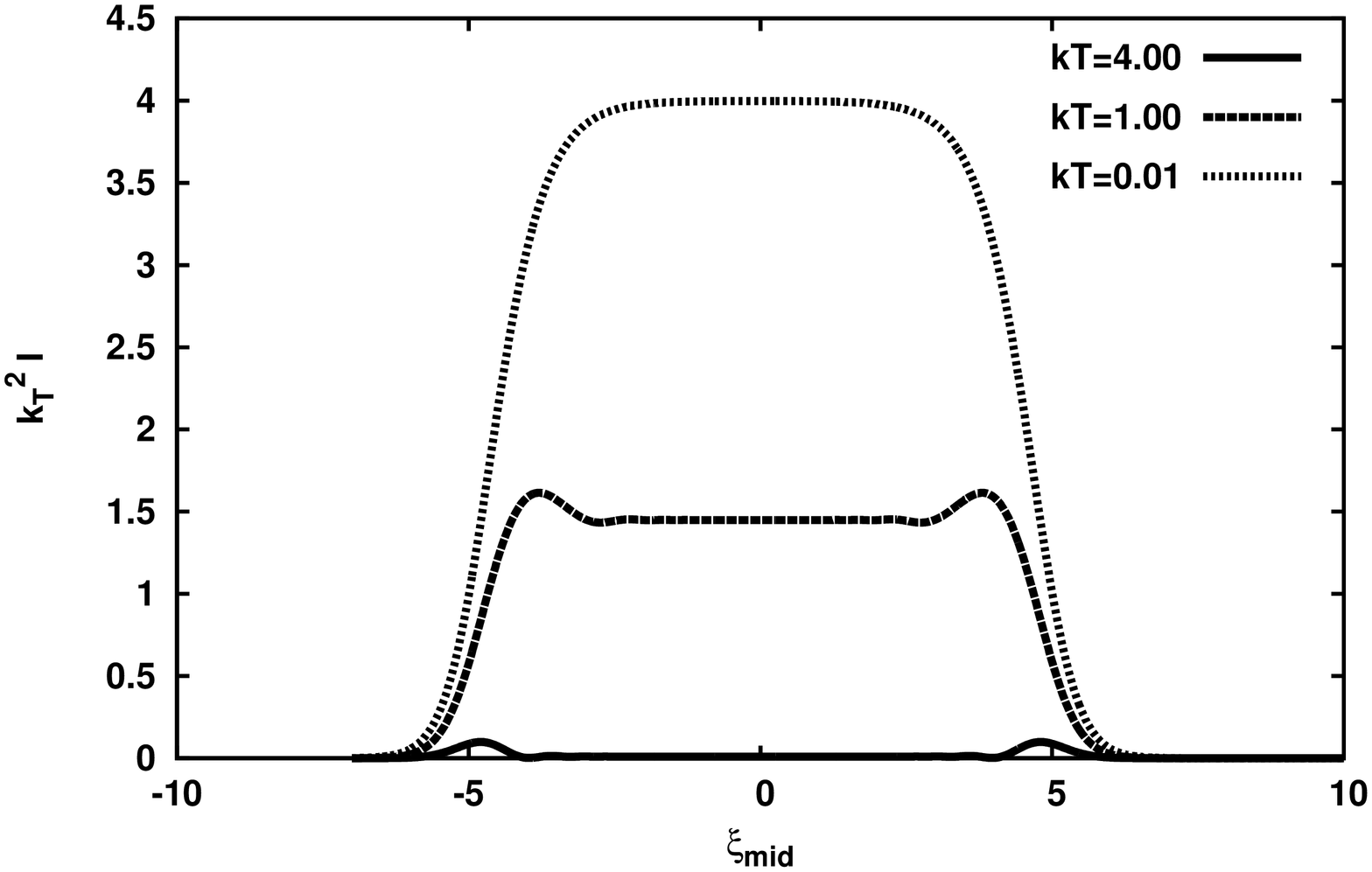}
\includegraphics[width=12pc,angle=-90]{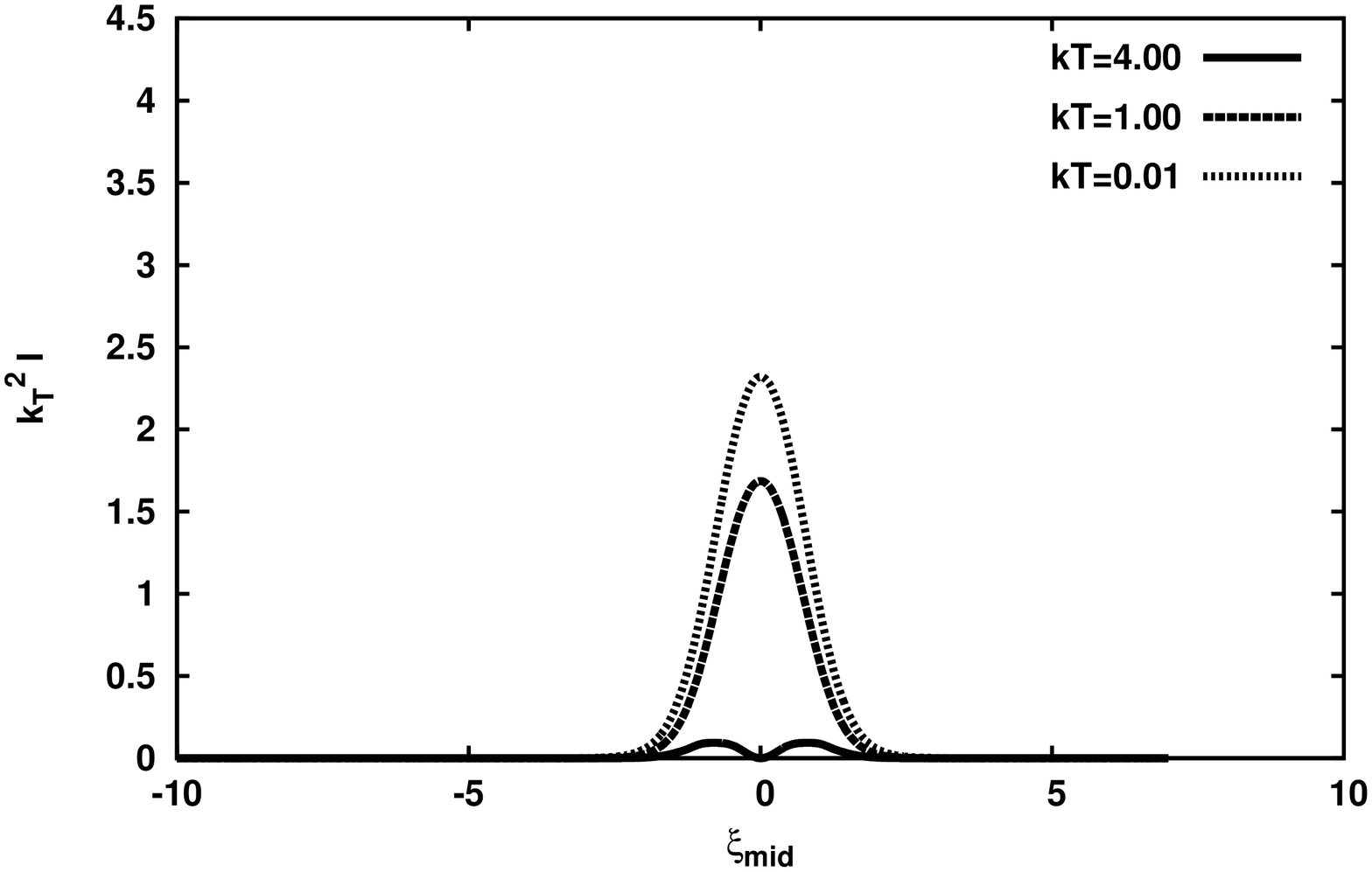}
\caption{\label{UnruhLong} Numerically calculated differential photon rapidity distribution from a
classical point charge moving straight with constant deceleration for a finite time. Left for
long, right for short deceleration window.}
\end{figure}

In our experiments, however, we cannot suppose monochromatic radiators or
movements with forever constant acceleration. Towards obtaining semiclassical
photon spectra from more realistic scenarios work is in progress\cite{WorkInProgress}.
As a preliminary a few characteristic photon rapidity distributions are shown
in Figure \ref{UnruhLong} at different transverse momenta, $k_T$, of the photon.
While long deceleration times lead to a flat rapidity distribution, like the Bjorken-flow
scenario, pictures produced using short times rather resemble those obtained by the Landau-flow scenario.
Not only a pseudo-temperature, but also a pseudo-flow occurs in these patterns.

\noindent

Hydrodynamics is well and alive among the theories in our field embracing a list
of intriguing questions to be clarified. A list of talks, like
{	
	Effects of Jets in the Flow Observables (Takahashi),
	Thermal Equilibrium and ... Initial Condition (Kodama),
	Landau hydrodynamics... (Tamosiunas),
	Shear viscosity of hadrons... (Wiranata),
	Turbulence, Vorticity and Lambda Polarization (Csernai),
	Hydrodynamic models of particle production (Bozek),
	QGP viscosity and the flow ... (Song),
	Event-by-event correlation... (Molnar),
	Elliptic Flow from CGC (Scradina),
	Dynamical freeze-out in event-by-event hydro... (Huovinen),
} more than proves it.

\section{Hadronization}

\vs
The main question of the hadronization process is since years whether its result
is phase space dominated (and in this case statistical descriptions are relevant)
or elmentary QCD matrix elements play the main role (and in this case only a
microdynamical approach prevails). Theoretically both factors can be studied.

We had talks among others on Monte Carlo (Werner), 
Strongly interacting parton-hadron matter... (Bratkovskaya), 
Gluon radiation by heavy quarks (Gousset),
Diffusion of Non-Gaussianity ... (Kitazawa), 
Strangeness baryon to meson ratio (Flores), 
on the Excluded-Volume Model (Tiwari), 
on  Interpretation of strange hadron production at LHC (Petran),
and on Chemical freezeout via HRG (Chatterjee). 

Is a ''thermal model'' or a ''statistical model''
is a good approximation to the complicated quantum (chromo-) dynamics?
By seeking an answer it can be helpful to factorize our search into several steps.
Such a factorizing of the ''thermo'' concept should include 
i) finding of a scaling variable which unifies different mass hadron $p_{\perp}$-spectra,
ii) testing the quark coalescence hypothesis, in case baryon and meson branches differ after the above step,
iii) finding out the unique functional form on the quark (parton) level describing all hadron spectra,
iv) testing trends with binary scaling, participant number, rapidity window, etc.
v) and only in a final step interpreting the parameters 
in physical terms (as on/off-equilibrium, finite/infinite size,
quantum/classical, phase space/matrix element dominated).
This defactorization, shown on a simple example what we did on RHIC data\cite{APOLONB}, 
involves step by step the following questions:
 \begin{enumerate}
  \item 
	  Is it   $f(p_T,m)=f(E_v(p_t,m)-\mu(m))=f(X)$ ?
  
   \item Is there a quark scaling? Is it $f_h(X) = f_q^n(X/n)$  for $n=2,3$?

   \item Do we see an (X-)energy distribution? 
	  $f(X) \sim \left(1+a \beta X \right)^{-1/a} \: {\longrightarrow} \: e^{-\beta X}$

   \item Are there trends with ($N_{part}, N_{bin}, P(N), \sqrt{s}, \Delta\eta$ ...) ?
  
   \item How to interpret the parameters  $\beta, v, a=(q-1), \ldots$? 
 \end{enumerate}

Also thermal cosmology was  introduced by {Johann Rafelski}.
We cite his remark: 
''When old people make a new theory, they know all about assumptions, approximations,
implied or explicit. But young people, who learn it from a textbook, believe
  that this were the \quad{TRUTH}''. 
We agree,  scientists always should be careful.

\section{Quarks off and on Lattices}

\vs
We heard a number of talks about Quarkonia and Energy Loss, like
Heavy Quark Energy Loss (Horowitz),
... Boltzmann vs Langevin (Das),
Towards... QGP (Berrehrah),
Heavy vs light flavor energy loss...  (Uphoff),
Jet quenching and Heavy Quarks (Renk),
and about thermal behavior of quark matter, such as 
Free energy vs internal energy potential...  (Lee),
Quantum and semiclassical...  (Katz),
... finite magnetic mass...  (Djordjevic).
One of the most interesting questions was whether  AdS/CFT does well? 
From Thorsten Trenk's review the answer is ''not quite'', more precisely
''clear no for both light and heavy quarks! AdS techniques predict
too much suppression at LHC when tuned to RHIC and extrapolated.''

So it remains to study quark matter with the more traditional 
lattice regularization of QCD.
Comparisons between Lattice QCD results with other (non-AdS) models
were presented in talks about high-density quark matter (Torrieri),
Sphalerons (Chao), flavor hierarchy  (Bluhm), the PNJL model (Yamazaki)
or the original Nambu-Jona-Lasinio model for SU(3)f (Marty).
Full blood lattice results were reported in talks about
First Principles Calculation (Allton),
Role of fluctuations in detecting the QCD phase transition (Redlich),
and "ab initio Lattice QCD calculations" (Schmidt),
while alternative approaches as
a talk on Holographic descriptions of dense quark matter (Kumar) or
on chiral fluid dynamics (Herold) tried to keep balance.

\begin{figure}[H]
\centerline{
	\includegraphics[width=0.4\textwidth]{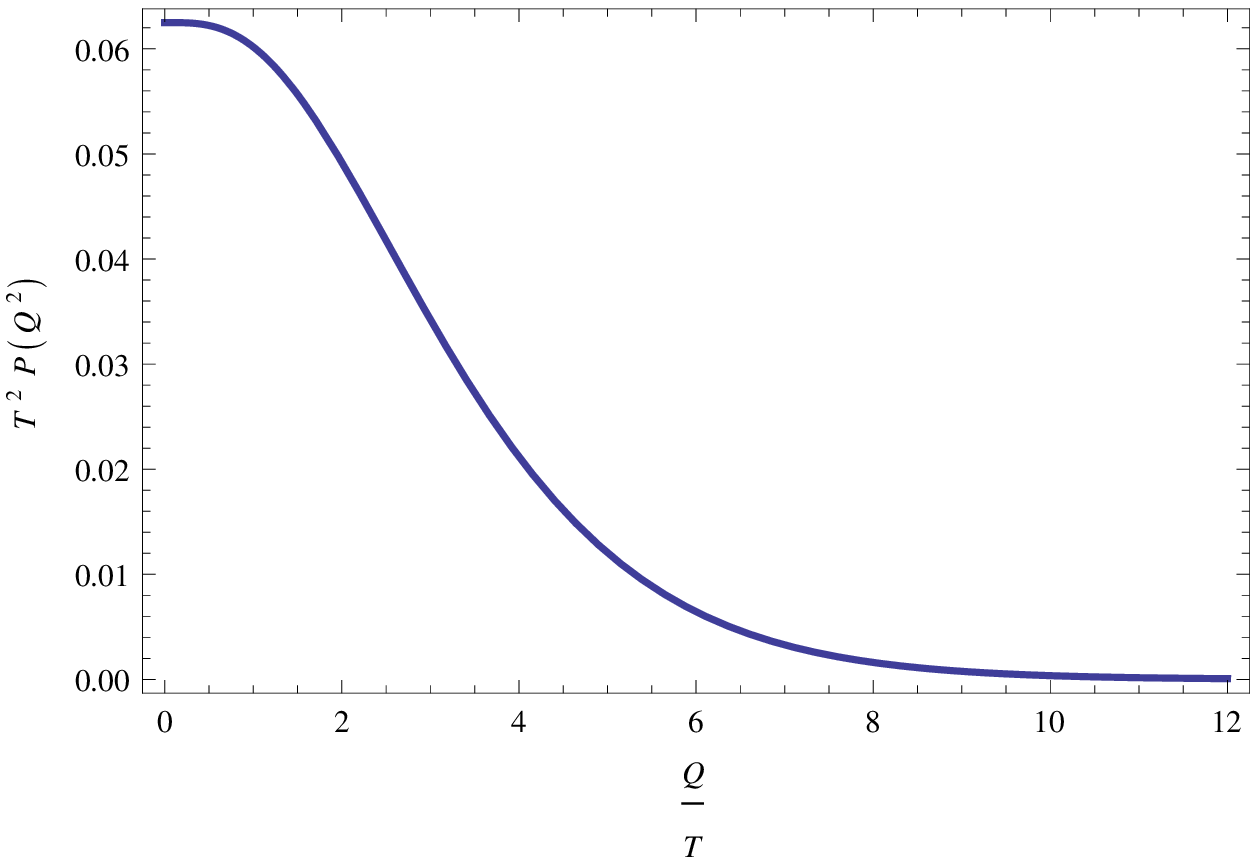} \qquad
	\includegraphics[width=0.4\textwidth]{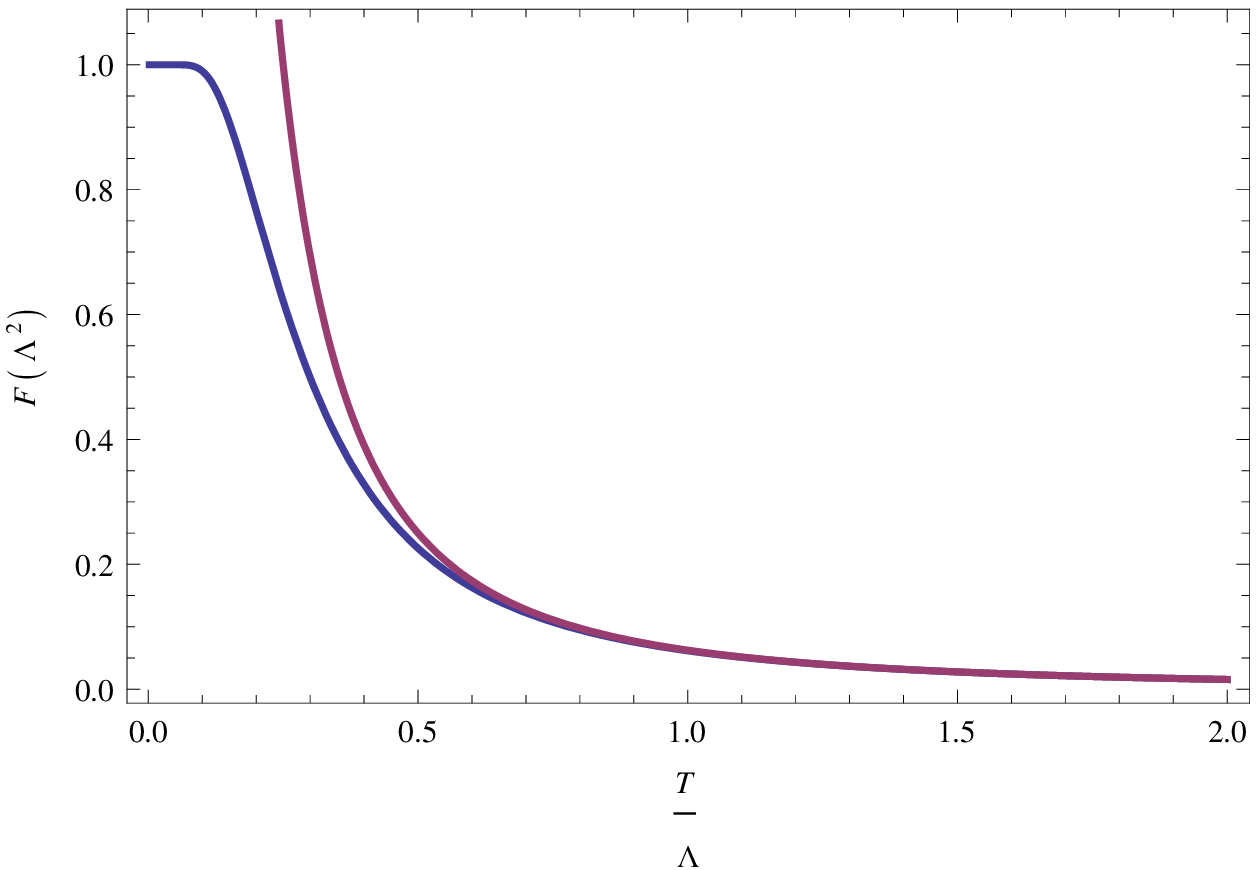}
}
\caption{\label{PQ2} Thermal probability to have $Q^2$ four-momentum squared (left)
and the integrated probability of being soft as a function of temperature (right)
for a pair of massless partons.}
\end{figure}

\vs
Most of these approaches were presented in a high-temperature context.
Therefore it is the right place to demonstrate that contrary to
a wide-spread false belief, high-T {\em does not mean} 
the unrestricted applicability of perturbative QCD (pQCD).
We may consider the following simple example.
Since pQCD relies on high $Q^2$ physics, we consider the
thermal distribution of $Q^2$ in an ideal Boltzmann gas of massless partons:
\be
P(Q^2)=\frac{\iiint\!dE_1 \,dE_2 \,d\cos\theta \, \: E_1^2 E_2^2 \, e^{- \beta (E_1+E_2) } 
\, \delta \left(Q^2 - 2 E_1 E_2(1 - cos\theta)\right)}
{\iiint\!dE_1 \,dE_2 \,d\cos\theta \, \: E_1^2 E_2^2 \, e^{- \beta (E_1+E_2) }}.
\ee{thermalQ2}
This integrals can be analytically solved, the result contains some Bessel K-functions:
\be
P(Q^2)=\frac{1}{64 T^2} \left(\frac{Q^3}{T^3} K_1 \left(\frac{Q}{T}\right)+ 
	2 \left(\frac{Q^2}{T^2}\right) K_2 \left(\frac{Q}{T}\right)\right)
\ee{thermalQ2_RESULT}
This Boltzmann-Gibbs $Q^2$ distribution is shown in Figure \ref{PQ2} (left).
While the width of the distribution does scale with $T$, the maximum probability is
at $Q=0$ at any temperature. At arbitrary high temperature non-perturbative effects
prevail, only their effect may diminish in some selected (infrared safe) observables.

Regarding an order parameter of Non-Perturbativity, the
thermal expectation value of ''being soft'' may be defined as follows:
\be
F(\Lambda^2) := \langle \theta ( \Lambda^2-Q^2) \rangle = \int_0^{\Lambda^2} P(Q^2) \,dQ^2.
\ee{thermalINT}
This definition selects out the integrated distribution function, pictured
in Figure \ref{PQ2} (right). The upper line belongs to the
perturbative approximation yielding $\Lambda^2/16T^2$. 
While the full result is saturated until about $T\approx\Lambda/6$,
which can be considered as $T_c\approx 167$ MeV, this approximation starts to be really good
only at $T \approx \Lambda$.
Hadronic effects may therefore well prevail until $T \approx 1$ GeV.
Between $T_1 \approx 1/6 $ GeV and $T_2 \approx 1$ GeV  both pQCD and non-pQCD worlds reign.

\section{Highlights}

\vs
Finally let us summarize theory highlights of this meeting. In our view
either new theories or new aspects, or new proposals for measurable effects,
perhaps new calculations on hard old problems are worth considering as
highlights, those which give or promise new insights into (collective or
individual, strange or less-strange) hadronic physics.

Towards a possibly new theory align efforts to derive the
EoS, the transport coefficients, heat capacities (acting on the
Tsallis parameter via the formula q=1+1/C), the distribution of fluctuations.
Expectations and hopes are related to
the determination of properties of elementary matter, to
comparison of phenomenology based findings with Lattice QCD calculational results,
to the very determination of an initial state just after the collision,
and to a deepened theoretical understanding of the birth of hadrons.

\vs
A new aspect connected with a new proposal was peresented by Becattini 
and Csernai\cite{BECATTINI,CSERNAI}
about the $\Lambda$-polarisation, enriching the hydrodynamical
approach with vorticity and spin effects.
New calculations seem to bring new conclusions: different confinement
temperature for the strange than for the non-strange sector -- as it has been
cited in Markus Bleicher's opening talk on theory news.
Certainly a new insight was expressed in the analysis of Klaus Werner
about the EPOS simulation, namely that Flow is Everywhere...
Another, relatively new proposal was mediated by Krzysztof Redlich: 
Fluctuations carry the most important message about quark matter and
about the phase transition. He found the so far most beautiful order parameter,
jumping from one value to another really sharply at $T_c$, in form of a 
sophisticated ratio of susceptibilities.
The proposal by Giorgio Torrieri to seek for quarkyonic signals in
the $v_2$ coefficient against $p_T$ plot while keeping the fluctuations
was also a novelty.

Finally other topics, like the astronuclear approach
in Alford's review taught us that some theories do not survive
observations. This clears ground for optimism.

In summary, quoting a commercial text discovered on a bottle of cider,
just replacing the word ''apple'' by theory, and the word ''ice''
by (experimental) ''data'' and ''Bulmers original cider'' by 
''Our original quark matter'', comes as follows:
{\em We use big theories, small theories, juicy theories and bittersweet
theories to make the well-balanced and medium-sweet flavor
of our original quark matter that you know and love.
Enjoy poured over data for ultimate refreshment.} 

\vs

{\bf Acknowledgement} \quad This work has been supported by the Hungarian National
Rsearch Fund (OTKA K104260) and by the Helmholtz International Center for FAIR
within the framework of the LOEWE program launched by the State of Hesse.

\end{document}